\documentclass{moriond}
\usepackage{amsmath,mathtools}

\newcommand{\Photo}{}
\newcommand{\rGW}{\rho_\text{GW}}
\newcommand{\rp}{\rho_\phi}
\newcommand{\rR}{\rho_R}
\newcommand{\Eom}{E_\omega}
\newcommand{\Trh}{T_\text{rh}}
\newcommand{\Tmax}{T_\text{max}}
\newcommand{\arh}{a_\text{rh}}
\newcommand{\mrh}{m_\text{rh}}
\newcommand{\oGW}{\Omega_\text{GW}}
\newcommand{\gs}{g_\star}
\newcommand{\gss}{g_{\star s}}

\begin{document}
\vspace*{4cm}
\title{Probing  Reheating with Gravitational Waves from Graviton Bremsstrahlung}
\author{Basabendu Barman}
\address{Department of Physics, School of Engineering and Sciences, SRM University-AP\\
Amaravati 522240, India}
\author{Nicol\'as Bernal~\footnote{Speaker}}
\address{New York University Abu Dhabi\\
PO Box 129188, Saadiyat Island, Abu Dhabi, United Arab Emirates}
\author{Simon Cléry, Yann Mambrini}
\address{Universit\'e Paris-Saclay, CNRS/IN2P3, IJCLab, 91405 Orsay, France}
\author{Yong Xu}
\address{\it PRISMA$^+$ Cluster of Excellence and Mainz Institute for Theoretical Physics\\
Johannes Gutenberg University, 55099 Mainz, Germany}
\author{Óscar Zapata}
\address{Instituto de Física, Universidad de Antioquia\\
Calle 70 \# 52-21, Apartado Aéreo 1226, Medellín, Colombia}

\maketitle
\abstracts{In this talk, based on Refs.~\cite{Barman:2023ymn,Barman:2023rpg,Bernal:2023wus}, we discuss the production of primordial gravitational waves (GW) sourced by graviton bremsstrahlung during inflationary reheating. For reheating, we consider inflaton decays and annihilations into pairs of bosons or fermions, assuming an inflaton $\phi$ that oscillates around a generic monomial potential $V(\phi) \propto \phi^n$. The GW spectrum exhibits distinct features depending on the underlying reheating dynamics, which is controlled by the inflaton potential and the type of coupling between the inflaton and the matter fields. We show that the produced stochastic GW background could be probed in next-generation GW detectors, especially at high frequencies. We further highlight the potential of bremsstrahlung-induced GW to probe the underlying dynamics of reheating.}

%%%%%%%%%%%%%%%%%%%%%%%%%%%%%%%%%%%%%
\section{Introduction}
%%%%%%%%%%%%%%%%%%%%%%%%%%%%%%%%%%%%%
Cosmic inflation is an elegant paradigm for resolving several problems in cosmology.~\cite{Lyth:2009zz} The simplest scenario is single-field inflation driven by the vacuum energy of a scalar field $\phi$, leading to a quasi-exponential expansion of the universe. Consequently, the universe becomes extremely cold at the end of inflation because of the exponential dilution. However, the success of Big Bang nucleosynthesis requires a thermal background as the initial condition. Understanding the transition from the cold state at the end of inflation to a hot and thermal universe is crucial. Reheating was proposed to explain the transition to the radiation-dominated epoch. The simplest idea of reheating is to consider couplings between inflaton and lighter daughter particles while the inflaton oscillates around the minimum of its potential. The detailed reheating dynamics is controlled by the shape of the potential of the inflaton and the couplings to matter. Due to the unavoidable coupling between the metric and the energy-momentum tensor $T_{\mu\nu}$, gravitons $h_{\mu\nu}$ (understood as perturbations of the metric $g_{\mu \nu}$ around a flat spacetime $\eta_{\mu \nu}$ as $g_{\mu \nu} \simeq \eta_{\mu \nu} + \frac{2}{M_P}\, h_{\mu \nu}$) are sourced by the vertex~\cite{Choi:1994ax}
\begin{equation} \label{eq:int2} 
    \sqrt{-g}\, \mathcal{L}_{\rm int} \supset -\frac{1}{M_P}\, h{\mu \nu}\, T^{\mu \nu}\,,
\end{equation} 
where $M_P \simeq 2.4 \times 10^{18}$ GeV is the reduced Planck mass. After production, these gravitons propagate throughout the universe, giving rise to a homogeneous and isotropic stochastic gravitational wave (GW) background.~\cite{Barman:2023ymn,Barman:2023rpg,Bernal:2023wus,Nakayama:2018ptw,Huang:2019lgd} We demonstrate that such bremsstrahlung-induced GWs can be used to probe the underlying reheating dynamics.

%%%%%%%%%%%%%%%%%%%%%%%%%%%%%%%%%%%%%
\section{Inflationary Reheating}
%%%%%%%%%%%%%%%%%%%%%%%%%%%%%%%%%%%%%
We assume inflaton $\phi$ interact with a real scalar $\varphi$ and a vector-like Dirac fermion $\psi$ through the Lagrangian density
\begin{equation}\label{eq:couplings}
    \mathcal{L} \supset \mu\, \phi\, \varphi^2 + \sigma\, \phi^2\, \varphi^2  + y\, \bar\psi \psi\, \phi\,,
\end{equation}
with $\mu$ being a mass dimension, while $\sigma$ and $y$ are dimensionless couplings. These interactions give rise to 1-to-2 and 2-to-2 inflaton decays and annihilations for reheating.

Additionally, we consider the post-inflationary oscillation of the inflaton $\phi$ at the bottom of a monomial potential $V(\phi)$ of the form
\begin{equation} \label{eq:inf-pot}
    V(\phi) = \lambda\, \frac{\phi^n}{M_P^{n - 4}}\,,
\end{equation}
where $\lambda$ is a dimensionless coupling. The effective mass $m(a)$ for the inflaton can be obtained from the second derivative of its potential and reads
\begin{equation} \label{eq:inf-mass1}
    m^2(a) \equiv \frac{d^2V}{d\phi^2} = n\, (n - 1)\, \lambda\, \frac{\phi^{n - 2}}{M_P^{n - 4}}
    \simeq n\, (n-1)\, \lambda^\frac{2}{n}\, M_P^\frac{2\, (4 - n)}{n} \rp(a)^{\frac{n-2}{n}},
\end{equation}
as a function of the cosmic scale factor $a$, and in terms of the inflaton energy density $\rp$. The evolution of $\rp$ is given by
\begin{equation} \label{eq:drhodt}
    \frac{d\rp}{dt} + \frac{6\, n}{2 + n}\, H\, \rp = - \frac{2\, n}{2 + n}\, \Gamma^{(0)}\, \rp\,,
\end{equation}
where $H$ is the Hubble expansion rate and $\Gamma^{(0)}(a)$ the 2-body decay or annihilation rate for inflatons into SM states. We note that the inflaton energy density takes the form
\begin{equation}
    \rp(a) \propto a^{-\frac{6\, n}{n+2}}
\end{equation}
in the regime $\Gamma^{(0)} \ll H$  during reheating.
Inflaton decays and annihilations heat the universe, producing the SM bath. The evolution of the SM radiation energy density $\rR$ is governed by the Boltzmann equation
\begin{equation} \label{eq:rR}
    \frac{d\rR}{dt} + 4\, H\, \rR = + \frac{2\, n}{2 + n}\, \Gamma^{(0)}\, \rp\,.
\end{equation}
The expressions~\eqref{eq:drhodt} and~\eqref{eq:rR} are coupled through the Friedmann equation $ H^2 = (\rR + \rp) /(3\, M_P^2)$.
The SM thermal bath can be described by the photon temperature $T$ as $\rR(T) = \pi^2/30\, \gs\, T^4$, where $\gs(T)$ corresponds to the number of relativistic degrees of freedom contributing to the SM energy density. The end of the reheating period is defined as the moment at which SM radiation starts to dominate the total energy density of the universe and corresponds to a reheating temperature $\Trh$, or equivalently, to a scale factor $\arh \equiv a(\Trh)$. It is important to note that away from the instantaneous decay approximation of the inflaton, the SM bath could reach temperatures higher than $\Trh$, up to $T = \Tmax$. Finally, the variable inflaton mass in Eq.~\eqref{eq:inf-mass1} can conveniently be rewritten as
\begin{equation} \label{eq:inf-mass}
    m(a) = \mrh \left(\frac{\arh}{a}\right)^\frac{3\, (n - 2)}{n + 2},
\end{equation}
where $\mrh \equiv m(\arh)$. It is interesting to note that for non-quadratic potentials (that is $n \neq 2$), $m$ has a field dependence that, in turn, leads to an inflaton decay or annihilation rate with a scale factor dependence.

The evolution of the SM energy density strongly depends on the characteristic of the interaction rate $\Gamma^{(0)}$: the nature of the process (decay or annihilation) and the spin of the final-state particles. From Eq.~\eqref{eq:couplings}, one has
\begin{equation}\label{eq:rates}
    \Gamma^{(0)}(a) =
    \begin{dcases}
        \frac{y_\text{eff}^2}{8 \pi}\, m(a) &\text{ for decays into fermions,}\\
        \frac{1}{8 \pi}\, \frac{\mu_\text{eff}^2}{m(a)} &\text{ for decays into scalars,}\\
        \frac{\sigma_\text{eff}^2}{8 \pi}\, \frac{\rp(a)}{m^3(a)} &\text{ for annihilations into scalars,}
    \end{dcases}
\end{equation}
for massless final-state particles. Contact interactions between two inflatons and two fermions are not allowed in a renormalizable framework and are therefore disregarded. The couplings $y_\text{eff}$, $\mu_\text{eff}$ and $\sigma_\text{eff}$ are effective in the sense that they are averaged over inflaton oscillations.~\cite{Ichikawa:2008ne} From Eq.~\eqref{eq:rates} it follows that during reheating the SM temperature scales as
\begin{equation}
    T(a) \simeq \Trh \left(\frac{\arh}{a}\right)^\alpha,
\end{equation}
with $\alpha = \frac{3\, (n-1)}{2\, (n+2)}$ for decays into fermions, $\alpha = \frac{3}{2\, (n+2)}$ for decays into scalars, or $\alpha = \frac{9}{2\, (n+2)}$ for annihilations into scalars. For the latter case, if $n = 2$ (and more generally if $n < 5/2$), the SM radiation cannot overtake the inflaton energy density, which implies that reheating cannot proceed successfully.

%%%%%%%%%%%%%%%%%%%%%%%%%%%%%%%%%%%%%
\section{Primordial Gravitational Waves from Graviton Bremsstrahlung}
\begin{figure}[t!]
    \def\sepf{0.45}
	\centering
      \includegraphics[scale=\sepf]{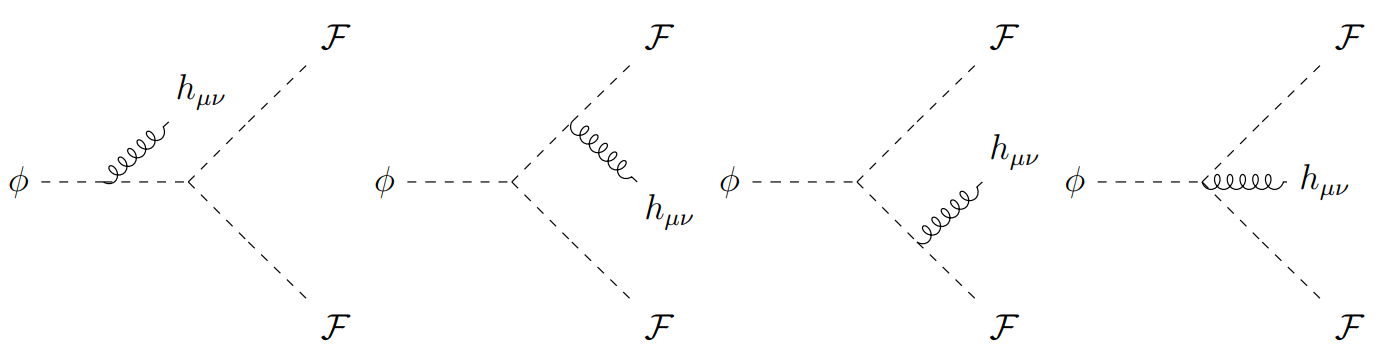}
    \caption{Inflaton decays into a pair of $\mathcal{F} \in \{\varphi, \psi\}$ and a graviton $h_{\mu\nu}$.} 
    \label{fig:3body}
\end{figure}

\begin{figure}[t!]
    \def\sepf{0.82}
    \centering
    \includegraphics[scale=\sepf]{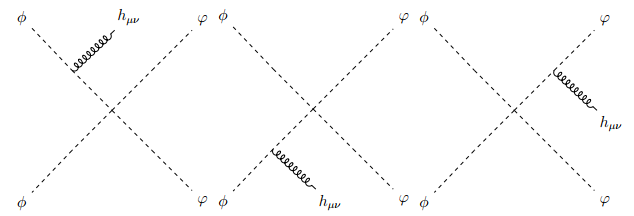}
    \includegraphics[scale=\sepf]{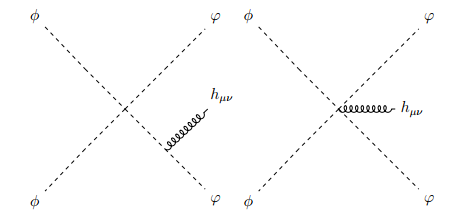}
    \caption{Inflaton annihilation into a pair of $\varphi$ and a graviton $h_{\mu\nu}$.}
    \label{fig:anni}
\end{figure}

%%%%%%%%%%%%%%%%%%%%%%%%%%%%%%%%%%%%%
Independently of the way reheating proceeds, gravitons are unavoidably produced by 3-body decays or annihilations of inflatons; the corresponding Feynman diagrams are shown in Figs.~\ref{fig:3body} and~\ref{fig:anni}. The differential rate $d\Gamma^{(1)}/d\Eom$ with respect to the graviton energy $\Eom$, in the case of massless final-state particles, is
\begin{equation}
    \frac{d\Gamma^{(1)}}{d\Eom} =
    \begin{dcases}
        \frac{y_\text{eff}^2}{64 \pi^3} \left(\frac{m}{M_P}\right)^2 \frac{(1-2x) \left[2x(x-1) + 1\right]}{x} &\text{ for decays into fermions,}\\
        \frac{1}{64\pi^3} \left(\frac{\mu_\text{eff}}{M_P}\right)^2 \frac{(1-2x)^2}{x} &\text{ for decays into scalars,}\\
        \frac{\sigma_\text{eff}^2}{8 \pi^3}\, \frac{\rp}{m^2\, M_P^2}\, \frac{(1 - x)^2}{x} &\text{ for annihilations into scalars,}
    \end{dcases}
\end{equation}
where $x \equiv \Eom/m$, with the kinematical threshold  $x < 1/2$ for decays or $x < 1$ for annihilations.
%%%%%%%%%%%%%%%%%%%%%%%%%%%%%%%%%%%%%%%%%%
%%%%%%%%%%%%%%%%%%%%%%%%%%%%%%%%%%%%%%%%%%
%%%%%%%%%%%%%%%%%%%%%%%%%%%%%%%%%%%%%%%%
%%%%%%%%%%%%%%%%%%%%%%%%%%%%%%%%%%%%%%%%

The evolution of the GW spectrum as a function of the scale factor $a$ depends on the ratio of energy densities stored in GWs $\rGW$ and in SM radiation, and can be tracked using the Boltzmann equation
\begin{equation} \label{eq:dGWdR2}
    \frac{d}{da} \frac{d(\rGW/\rR)}{d\Eom'} \simeq \frac{\arh}{a} \frac{2n}{2 + n} \frac{1}{a\, H(\Trh)} \left(\frac{\arh}{a}\right)^{\frac{3n}{2 + n} - 4 \alpha} \left[\frac{d\Gamma^{(1)}}{d\Eom'} \frac{\Eom'}{m} - \frac{a}{\arh} \frac{d(\rGW/\rR)}{d\Eom'} \Gamma^{(0)}\right],
\end{equation}
where $\Eom'$ correspond to the graviton energy redshifted to the end of reheating. The primordial GW spectrum $\oGW(f)$ at present per logarithmic frequency $f$ is
\begin{equation} \label{eq:oGW}
    \oGW(f) = \Omega_\gamma^{(0)}\, \frac{\gs(\Trh)}{\gs(T_0)} \left[\frac{\gss(T_0)}{\gss(\Trh)}\right]^{4/3}\,\frac{d(\rGW(\Trh)/\rR(\Trh))}{d\ln \Eom'}\,,
\end{equation}
where $\Omega_\gamma^{(0)} h^2 \simeq 2.47 \times 10^{-5}$ is the current photon density, $T_0 \simeq 2.73$~K is the CMB temperature, and $\gss(T)$ is the number of relativistic degrees of freedom that contribute to the SM entropy. Additionally, the present GW frequency corresponds to a graviton energy $\Eom'$ at the end of reheating through
\begin{equation}
    f = \frac{\Eom'}{2 \pi}\, \frac{\arh}{a_0} = \frac{\Eom'}{2 \pi}\, \frac{T_0}{\Trh} \left[\frac{\gss(T_0)}{\gss(\Trh)}\right]^{1/3},
\end{equation}
with $a_0$ being the scale factor at present. Note that the kinematical upper bound of the graviton energy is translated into an upper bound on the frequency at present given by
\begin{equation} \label{eq:fup}
    f \leq \frac{\beta\, \mrh}{4 \pi}\, \frac{T_0}{\Trh} \left[\frac{\gss(T_0)}{\gss(\Trh)}\right]^{1/3} \times
    \begin{dcases}
        1 &\text{ for } n \leq 4\,,\\
        \left(\frac{\Tmax}{\Trh}\right)^\frac{2\, (n-4)}{\alpha\, (n+2)} &\text{ for } n \geq 4\,,
    \end{dcases}
\end{equation}
with $\beta = 1$ for inflaton decays, while $\beta = 2$ for annihilations.

\begin{figure}[t!]
    \def\sepf{0.50}
    \centering
    \includegraphics[scale=\sepf]{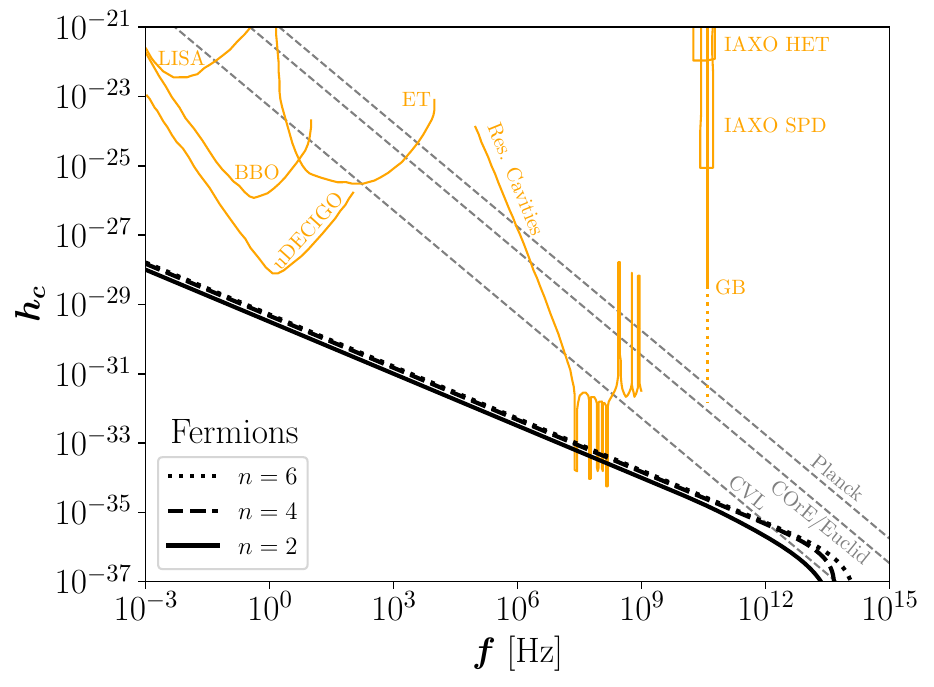}
    \includegraphics[scale=\sepf]{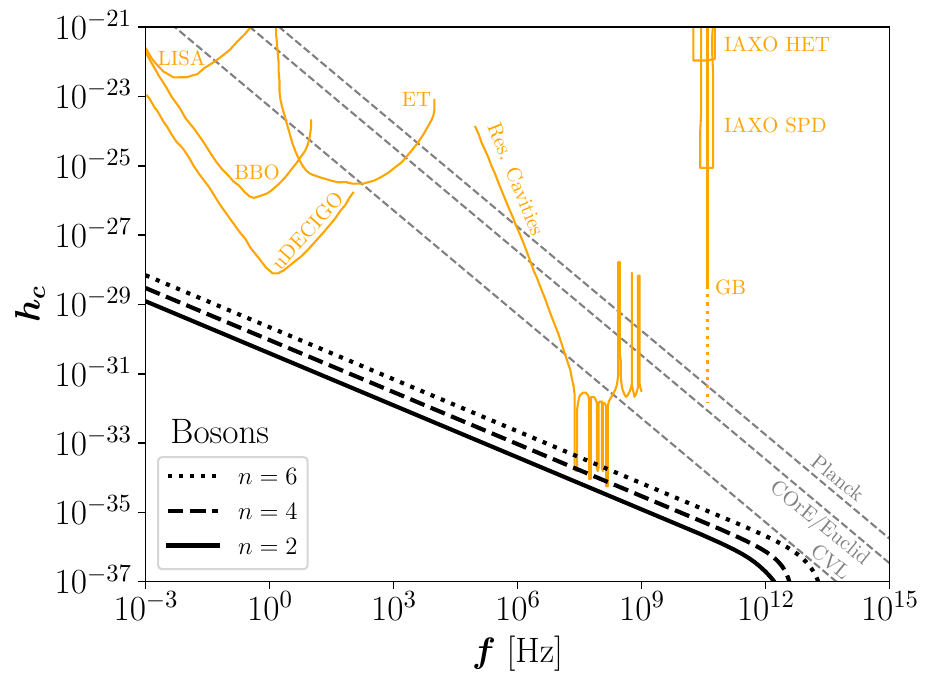}
    \caption{GW spectrum from inflaton decays as a function of frequency, considering fermionic reheating, with $\mrh = 5 \times 10^{16}$~GeV, $\Trh = 10^{13}$~GeV, and $\Tmax/\Trh = 10$ (left) or a bosonic reheating with $\mrh = 5 \times 10^{15}$~GeV, $\Trh = 10^{13}$~GeV, and $\Tmax/\Trh = 2$ (right). The sensitivities of future GW detectors are represented by solid orange lines.}
    \label{fig:GWs}
\end{figure}

In order to project limits from different GW experiments, it is convenient to construct the dimensionless strain $h_c$ in terms of the GW spectral energy density as
\begin{equation}
    h_c(f) = \frac{H_0}{f}\, \sqrt{\frac{3}{2 \pi^2}\, \oGW(f)} \simeq 1.26 \times 10^{-18} \left(\frac{\rm{Hz}}{f}\right) \sqrt{\oGW(f)\, h^2}\,,  
\end{equation}
where $H_0 \equiv H(T_0) \simeq 1.44 \times 10^{-42}$~GeV is the present-day Hubble rate and $h = 0.674$.~\cite{Planck:2018vyg} The left panel of Fig.~\ref{fig:GWs} shows the dimensionless strain of GW as a function of frequency, considering fermionic reheating for $\mrh = 5 \times 10^{16}$~GeV, $\Trh = 10^{13}$~GeV, and $\Tmax/\Trh = 10$. In this case, the increase in $n$ slightly affects the GW spectrum. If we go from $n = 2$ to $n = 4$, the spectrum is boosted by a factor $\frac34 (\Tmax/\Trh)^{8/9} / \ln(\Tmax/\Trh)$. Similarly, going from $n = 4$ to $n = 6$ one gets an $\mathcal{O}(1)$ factor $\sim \frac34 (\Tmax/\Trh)^{8/45}$. The right panel of Fig.~\ref{fig:GWs} corresponds to the case of bosonic decay, for $\mrh = 5 \times 10^{15}$~GeV, $\Trh = 10^{13}$~GeV and $\Tmax/\Trh = 2$. In contrast to the fermionic case, for bosons, the increase in $n$ has a significant impact on the GW spectrum. If we go from $n = 2$ to $n = 4$, the spectrum is boosted by a factor $\frac34 (\Tmax/\Trh)^{8/3} / \ln(\Tmax/\Trh)$. Similarly, going from $n = 4$ to $n = 6$ one gets a factor $\sim \frac34 (\Tmax/\Trh)^{8/3}$. The prominent enhancement on the GW spectrum in the bosonic decay case is due to the fact that the released entropy, hence the dilution effect, is suppressed; cf. Eq.~\eqref{eq:rates}. In Fig.~\ref{fig:GWs} we also project the limits from several proposed GW detectors; see also Refs.~\cite{Barman:2023ymn,Barman:2023rpg,Bernal:2023wus}. Interestingly, bremsstrahlung-induced GWs might be within the reach of resonant-cavity detectors in the high-frequency regime. At lower frequencies, they might be detected by the future space-based GW detectors, e.g., uDECIGO for certain model parameters.
%%%%%%%%%%%%%%%%%%%%%%%%%%%%%%%%%%%%%%%%%%%%%%%%%%%
%%%%%%%%%%%%%%%%%%%%%%%%%%%%%%%%%%%%%%%%%%%%%%%%%%%

%%%%%%%%%%%%%%%%%%%%%%%%%%%%%%%%%%%%%%%%%%%%%%%%%%%%%%%%%%%%
\begin{figure}
    \def\sepf{0.60}
    \centering
     \includegraphics[scale=\sepf]{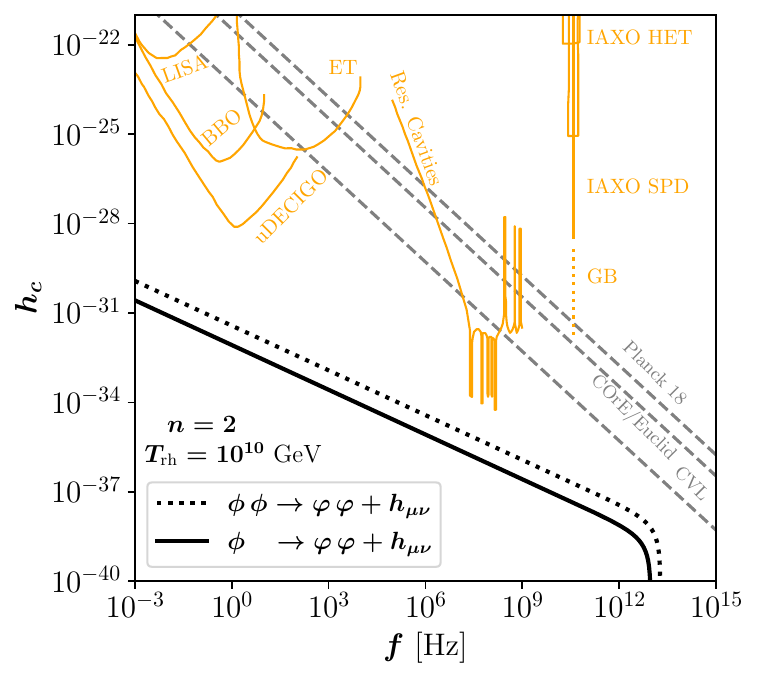}
     \includegraphics[scale=\sepf]{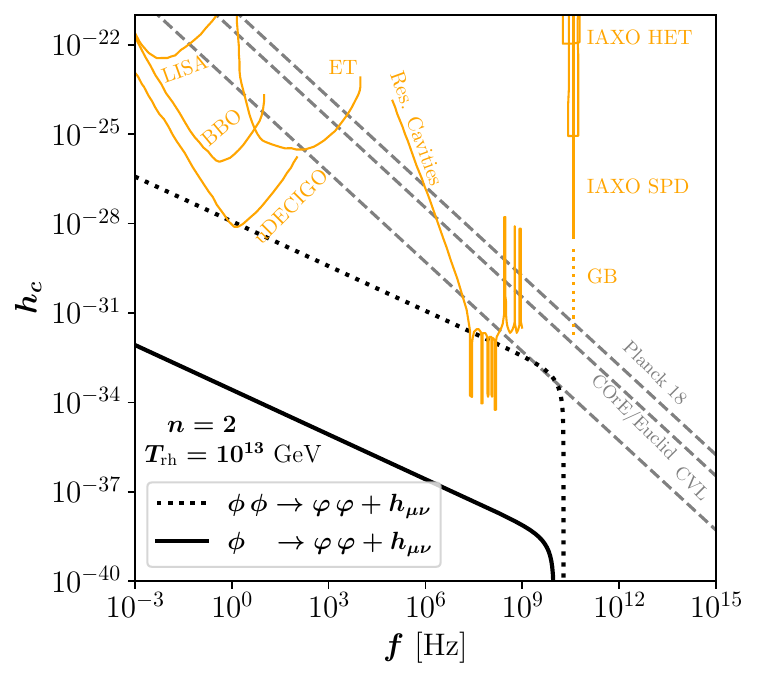}
    \caption{GW spectrum for the quadratic inflaton potential $n=2$ with $\Trh = 10^{10}$~GeV (left) and $\Trh = 10^{13}$~GeV (right). Solid black lines represent the signal for inflaton decays with $\mu=10^{10}$~GeV, whereas dashed black lines represent the signal for inflaton scattering with $\sigma = 1$. The sensitivities of future GW detectors are represented by solid orange lines.}
    \label{fig:GW_peak}
\end{figure}
%%%%%%%%%%%%%%%%%%%%%%%%%%%%%%%%%%%%%%%%%%%%%%%%%%%%%%%%%%%%
In the left panel of Fig.\ref{fig:GW_peak}, we illustrate the predicted GW spectrum for inflaton decay (solid black line) and scattering (dotted black line), assuming $\mu_\text{eff} = 10^{10}$~GeV, $\sigma_\text{eff} = 1$, $\Trh = 10^{10}$~GeV, and $n = 2$. For annihilations, the cutoff of the spectrum occurs at a frequency higher by a factor of two compared to the case of decays; see Eq.~\eqref{eq:fup}. Moving to the right panel of Fig.~\ref{fig:GW_peak}, corresponding to $\Trh=10^{13}$ GeV, we clearly observe that the spectrum derived from scattering, in the case $n=2$, can be larger by orders of magnitude than that generated by inflaton decay, especially for large $\Trh$. It is worth noting that this conclusion is based on the assumption that we do not restrict the reheating process to account for graviton bremsstrahlung. For $n>2$, one can assume that reheating via inflaton decay or annihilation also sources gravitons. In such cases, we find that the enhancement of the GW spectrum is more prominent in the decay case compared to the annihilation, mainly due to the suppression of the entropy dilution involved in the former.~\cite{Bernal:2023wus}

%%%%%%%%%%%%%%%%%%%%%%%%%%%%%%%%%%%%%5
\section{Conclusions}
%%%%%%%%%%%%%%%%%%%%%%%%%%%%%%%%%%%%%5
In this talk, we presented the inexorable production of stochastic gravitational waves (GWs) through graviton bremsstrahlung during reheating. It results from the unavoidable gravitational interaction arising from the graviton-matter coupling. We assumed an inflaton field $\phi$ oscillating around a monomial potential $\phi^n$, decaying or annihilating into pairs of bosons or fermions. We demonstrated that the GW spectrum depends on the inflaton potential shape parameter $n$ as well as its type of coupling to matter. Interestingly, for $n > 2$, there is a significant increase in GW amplitude during bosonic reheating. Also, for $n = 2$, the scattering of the inflaton can produce a much larger GW spectrum than the decay channel. These findings suggest that next-generation high-frequency GW experiments could illuminate the microscopic dynamics of reheating.

%%%%%%%%%%%%%%%%%%%%%%%%%%%%%%%%%%%%%5
\section*{Acknowledgments}
%%%%%%%%%%%%%%%%%%%%%%%%%%%%%%%%%%%%%5
NB received funding from the Spanish FEDER / MCIU-AEI under the grant FPA2017-84543-P. YX acknowledges the support from the Cluster of Excellence ``Precision Physics, Fundamental Interactions, and Structure of Matter'' (PRISMA$^+$ EXC 2118/1) within the German Excellence Strategy (Project No. 390831469).

%%%%%%%%%%%%%%%%%%%%%%%%%%%%%%%%%%%%%5
\section*{References}
%%%%%%%%%%%%%%%%%%%%%%%%%%%%%%%%%%%%%5

%%%%%%%%%%%%%%%%%%%%%%%%%%%%%%%%%%%%%5

\begin{thebibliography}{99}

\bibitem{Barman:2023ymn}
B.~Barman, N.~Bernal, Y.~Xu and \'O.~Zapata,
JCAP \textbf{05} (2023), 019
doi:10.1088/1475-7516/2023/05/019
[arXiv:2301.11345 [hep-ph]].

\bibitem{Barman:2023rpg}
B.~Barman, N.~Bernal, Y.~Xu and \'O.~Zapata,
Phys. Rev. D \textbf{108} (2023) no.8, 083524
doi:10.1103/PhysRevD.108.083524
[arXiv:2305.16388 [hep-ph]].

\bibitem{Bernal:2023wus}
N.~Bernal, S.~Cl\'ery, Y.~Mambrini and Y.~Xu,
JCAP \textbf{01} (2024), 065
doi:10.1088/1475-7516/2024/01/065
[arXiv:2311.12694 [hep-ph]].

\bibitem{Lyth:2009zz}
D.~H.~Lyth and A.~R.~Liddle,
``The primordial density perturbation: Cosmology, inflation and the origin of structure''

\bibitem{Choi:1994ax}
S.~Y.~Choi, J.~S.~Shim and H.~S.~Song,
Phys. Rev. D \textbf{51} (1995), 2751-2769
doi:10.1103/PhysRevD.51.2751
[arXiv:hep-th/9411092 [hep-th]].

\bibitem{Nakayama:2018ptw}
K.~Nakayama and Y.~Tang,
Phys. Lett. B \textbf{788} (2019), 341-346
doi:10.1016/j.physletb.2018.11.023
[arXiv:1810.04975 [hep-ph]].

\bibitem{Huang:2019lgd}
D.~Huang and L.~Yin,
Phys. Rev. D \textbf{100} (2019) no.4, 043538
doi:10.1103/PhysRevD.100.043538
[arXiv:1905.08510 [hep-ph]].

\bibitem{Ichikawa:2008ne}
K.~Ichikawa, T.~Suyama, T.~Takahashi and M.~Yamaguchi,
Phys. Rev. D \textbf{78} (2008), 063545
doi:10.1103/PhysRevD.78.063545
[arXiv:0807.3988 [astro-ph]].

\bibitem{Planck:2018vyg}
N.~Aghanim \textit{et al.} [Planck],
Astron. Astrophys. \textbf{641} (2020), A6
[erratum: Astron. Astrophys. \textbf{652} (2021), C4]
doi:10.1051/0004-6361/201833910
[arXiv:1807.06209 [astro-ph.CO]].

\end{thebibliography}
\end{document}